\PassOptionsToPackage{table}{xcolor}
\documentclass[acmsmall]{acmart}

\usepackage{graphicx}

\usepackage{colortbl}
\usepackage{tikz}

\usepackage[normalem]{ulem}
\useunder{\uline}{\ul}{}
\AtBeginDocument{%
  }

\setcopyright{cc}
\setcctype{by}
\acmJournal{PACMHCI}
\acmYear{2025} \acmVolume{9} \acmNumber{2} \acmArticle{CSCW121} \acmMonth{4} \acmPrice{}\acmDOI{10.1145/3711114}

\acmSubmissionID{V9cscw121}

\begin{document}

\title[Teen Discord Moderators]{``It's Great \textit{Because} It's Ran By Us'': Empowering Teen Volunteer Discord Moderators to Design Healthy and Engaging Youth-Led Online Communities}


\author{Jina Yoon}
\affiliation{%
  \institution{University of Washington}
  \city{Seattle}
  \country{USA}}
\email{jinayoon@cs.uw.edu}

\author{Amy X. Zhang}
\affiliation{%
  \institution{University of Washington}
  \city{Seattle}
  \country{USA}}
\email{axz@cs.uw.edu}

\author{Joseph Seering}
\affiliation{%
 \institution{KAIST}
 \city{Daejeon}
 \country{Republic of Korea}}
\email{seering@kaist.ac.kr}

\renewcommand{\shortauthors}{Jina Yoon, Amy X. Zhang, and Joseph Seering}

\begin{abstract}
Online communities can offer many benefits for youth including peer learning, cultural expression, and skill development. However, most HCI research on youth-focused online communities has centered communities developed by adults for youth rather than by the youth themselves. In this work, we interviewed 11 teenagers (ages 13-17) who moderate online Discord communities created \textit{by} youth, \textit{for} youth. Participants were identified by Discord platform staff as leaders of well-moderated servers through an intensive exam and application-based process. We also interviewed 2 young adults who volunteered as mentors of some of our teen participants. We present our findings about the benefits, motivations, and risks of teen-led online communities, as well as the role of external stakeholders of these youth spaces. We contextualize our work within the broader teen online safety landscape to provide recommendations to better support, encourage, and protect teen moderators and their online communities. This empirical work contributes one of the first studies to date with teen Discord moderators and aims to empower safe youth-led online communities.

\end{abstract}



\received{July 2024}
\received[revised]{October 2024}
\received[accepted]{December 2024}

\maketitle\section{Introduction}

Online communities offer a wide variety of benefits for youth, ranging from promotion of peer learning~\cite{jenkins2015participatory}, facilitation of diverse cultural expression~\cite{boyd2007why}, and support for the development of modern workplace skills~\cite{ito2013hanging}. Despite their potential, most HCI research on youth-focused online communities has focused on communities or programs developed by adults for youth, rather than by the youth themselves~\cite{jagannath2020talk,tekinbas2021designing,slovak2018mediating}. Prior work offers many clues for the potential value of youth-developed and youth-led communities. In early adolescence, individuals seek greater independence and autonomy~\cite{ryan2020intrinsic, deci2000and}, which youth-led online communities are well positioned to offer. Moreover, research has shown that early experiences participating in online communities can lead to an increase in appreciation for and participation in civic engagement, even if the community itself is not expressly political~\cite{kahne2013civic}. Within spaces such as these, recent research has highlighted how traditional, top-down moderation styles are less compatible with youth preferences in justice and harm resolution~\cite{schoenebeck2021youth}; allowing youth to lead their own communities could empower them to pursue approaches to justice that are more in line with their values.

However, there are many risks that might accompany the increased visibility and autonomy that come with the leadership of online communities. Platforms such as Discord and Roblox, which are very popular among youth users, have come under fire in the media for alleged prevalence of youth-specific harms. Moreover, prior work has highlighted the risks and burdens of community leadership for adult moderators. For example, Reddit moderators are more likely than average users to encounter harmful content, harassment, and emotional burnout~\cite{schopkegonzalez2022burnout,dosono2019moderation}. These issues have led scholars to critique such platforms for their reliance on unpaid labor~\cite{li2022measuring}, and these concerns would only grow more serious if many of the moderators were minors.

In this work, we explore the potential risks and benefits for greater youth involvement in the development and leadership of online communities as well as the opportunities for non-youth stakeholders to provide support. To do so, our site of study focuses on existing large youth-led communities ``in the wild'' on Discord that have grown organically. We partnered with Discord staff to recruit 11 teenagers (between ages 13-17) from the Moderator Mentorship Community (MMC), a program created by Discord to facilitate positive peer networks, create mentorship opportunities, and provide learning resources. To be admitted, participants had to already be a moderator of a large Discord server, pass the Discord Moderator Academy (DMA) exam, and be vetted by platform staff through a highly selective process. We also interviewed two young adults who volunteered as mentors in the MMC to get their perspectives as near-peer stakeholders. Our research questions were as follows:

\begin{itemize}
    \item \textbf{RQ1:} How do teen online community moderators benefit from leading and moderating their own communities?
    \item \textbf{RQ2:} What motivates teens to join and participate in youth-led online communities?
    \item \textbf{RQ3:} What role do external stakeholders play in supporting teens in online community leadership?
    \item \textbf{RQ4:} What risks do teens still face as online community moderators?
\end{itemize}

Our findings suggest that youth-led online communities, when supported well, can be beneficial to teen moderators and members \textit{as a direct result} of their peer-driven structure. Participants relayed that they gained valuable social and practical skills as a result of their experiences, and the authentic participation and leadership styles of peer leaders were well-aligned with youths' preferred models of governance, justice, and the cultures found in these communities. Some participants' communities even collaborated with official third-party businesses such as game developers, livestreamers, and influencers to develop thriving fan communities. These partnerships enabled benefits such as in-game rewards that attracted more members, but interactions with external organizations could also lead to complicated power dynamics, especially due to age differences.

There were many other stakeholders involved with participants' communities that helped keep make safe and successful. Participants shared that the MMC---the community they were recruited from---was one of the most important resources for their growth and development through peer discussion, dedicated mentorship, and access to platform support. However, despite being some of the most well-supported teens in the Discord safety ecosystem, our interviewees still face many risks intrinsic to voluntary community-based moderation. The most commonly discussed challenge was work-life balance, or difficulty setting boundaries on one's time and energy spent on managing their communities. Some participants also discussed encounters with harmful content and harassment. Additionally, despite having strong support systems online, participants did not always feel understood by members of their offline community such as their parents, guardians, and friends from school.

In our discussion, we contextualize our findings within the broader teen online trust and safety landscape to offer recommendations for stakeholders to better support teen moderators and their communities. We then conclude by revisiting what aspects of Discord's community-based and community-driven structures seemed to promote healthy and engaging teen-led spaces for our participants. In summary, we contribute this study of potential factors, benefits, stakeholders, and challenges that can result from empowering teen Discord moderators and their communities. This research is one of the first empirical studies to date with this important population and ultimately seeks to call attention to the value and need for alternative, peer-led social experiences for youth online.

\section{Related Work}
\subsection{Youth participatory culture on community-based platforms}
Online interest-based communities can be particularly beneficial for youth because they provide low barriers to self-expression, opportunities for informal mentorship, and a sense of social connection~\cite{ito2013hanging}. These factors are part of what scholars like Ito, Jenkins, and boyd have defined as \textit{participatory culture}~\cite{jenkins2015participatory}. These virtual spaces can be positive for youth because they promote peer-to-peer informal learning, diverse cultural expression, and development of modern workplace skills~\cite{ito2009living, fiesler2017growing, fields2014programming,brennan2010making, evans2017fanfiction}. Furthermore, platform affordances like pseudonymity and peer-matching can encourage more honest self-disclosure than conversations in-person, which can be crucial for youth seeking help seeking mental health support in private~\cite{bhattacharya2019teen, jin2023music, wadley2013participatory, fang2022peer}. These sociotechnical elements also make online communities valuable for marginalized youth such as LGBTQ+ teens exploring their identity~\cite{dym2019coming, mcinroy2022counter} or seeking publicly inaccessible health information~\cite{liang2020education}. 

These past works have surfaced valuable findings relevant to the present study, but they largely focus on either adult-led interventions~\cite{tekinbas2021designing,cramer2007everything} or small identity-based communities~\cite{kotut2022winds}. We differentiate our work in that we are interested in examining online Discord communities that are moderated by teens between 13-17 years old. Additionally, we aim to understand what factors might make these peer-led communities engaging and appealing at scale. Despite these known benefits, community-based platforms like Reddit, Twitch, and Discord are still much less visible in youth culture than massive network-based sites like TikTok and Instagram~\cite{gottfriend2023teens}, which have been more frequently criticized for potentially harmful design patterns such as infinite scrolling and short-form video~\cite{schellewald_understanding_2023, tabachnick_nebraska_2024}. Thus, one of the motivations of our work is to identify what aspects of teen-led online communities might make them appealing as alternative, healthy social experiences for youth. To answer this, we investigate the practices and values of large teen-led communities that are already successful ``in the wild'' today. 

\subsection{Challenges with voluntary online community-driven moderation}
\textit{Online volunteer community moderators} are users who voluntarily manage, build, and engage with online communities on platforms such as Reddit, Discord, Facebook Groups, and Twitch~\cite{seering2019moderator}. Volunteer moderators have been a part of the online social ecosystem since its origins in electronic bulletin board systems~\cite{stone1993vampires}, and the processes of volunteer moderation have been explored in depth in CSCW literature (e.g.,~\cite{seering2020reconsidering, gilbert2020cesspool, srinivasan2019removal, geiger2010work, seering2017shaping, chandrasekharan2018internet, jiang2019voice, kiene2019frames, jhaver2019explanations, chandrasekharan2019crossmod, cullen2022practicing, cai2021profiling, schluger2022proactive}). Per Seering et al.~\cite{seering2019moderator}, volunteer moderators engage in a variety of activities, ranging from proactively monitoring and participating in their communities to removing offending content and users to discussing and writing policies. While considerable public discussion surrounding content moderation has focused on content and user removal, much of what volunteer moderators do is actually focused on community-building~\cite{seering2022whomoderates, seering2022metaphors, gibson2023teams}. Volunteer moderators welcome new members, educate them about rules, participate in and encourage high-quality discussion, and may even organize community events or social activities. Teams of volunteer moderators for larger communities may be administratively sophisticated~\cite{gilbert2020cesspool} and employ a wide variety of technical tools to assist with their various duties~\cite{kiene2019frames}. Platforms like Discord, Reddit, and Twitch arguably could not exist without these volunteers, yet they remain largely unpaid, underappreciated, and face greater risks of harm online than average users due to their visibility~\cite{seering2022pride,gilbert2023towards}. Another common concern is that platforms may be exploiting the passion of volunteer moderators for free labor, which can result in burnout due to the sheer volume of interpersonal conflicts they mediate in their communities~\cite{seering2019beyond,yu2020fruit,schopkegonzalez2022burnout}. 

These concerns become even more serious when the volunteers in question are teens with developmental sensitivities \cite{orben2022windows}. Today, platforms including Discord, Reddit, Twitch, and Facebook, allow users as young as 13 years old to volunteer as a moderator in any online community. While the actual proportion of moderators under the age of 18 is not publicly known, many online communities related to topics popular among youth such as gaming, high school, or college prep are likely moderated by teens today. To our knowledge, there are no studies about this population to date. In this work, we seek to bridge this gap to ensure that teen moderators are protected against the potential risks and challenges of their online leadership positions. Additionally, even though community-based moderation raises many ethical tensions, it also provides many benefits for the volunteers and members including a sense of fulfillment, belonging, and community~\cite{seering2022pride}. To make this relationship more sustainable, researchers have proposed solutions like monetary compensation or social recognition, but these are largely based on studies with adults ~\cite{dosono2019moderation,li2022measuring, chang2011social}. We are concerned with the fact that users under the age of 18 may not be eligible for options like payment. Thus, in addition to preventing risks, we also will examine proactive ways to better compensate and support teen moderators.

\subsection{The role of platform governance and moderation on youth social media experiences}
Platform governance and moderation structures can greatly impact users' social media experiences~\cite{zhang_form_2024}. In recent years, researchers have experimented with multi-level governance structures to decentralize platform power and offer more modularity, agency, and control for users at large~\cite{jhaver_decentralizing_2023}. The impact and potential of these structural factors, however, remain highly understudied for youth social media experiences. Davis explored at a conceptual level how community-supported and self-directed experiences can benefit youth and adolescent development~\cite{davis2023technologys}. Landesman et al. have investigated how teens' moment-by-moment emotional states and mindfulness can influence their experiences on social media~\cite{landesman2024care}. 
Others have examined various features, affordances, and interactions that influence youth social media experiences, but these also generally focus on interactive aspects of more SNS-based platforms like content permanence and quantifiability~\cite{nesi_transformation_2018,nesi_transformation_2018_2}. 

There is an urgent need and opportunity to more closely examine how governance-related structures, and especially community-driven models, can benefit youth social media experiences. In this work, we seek to answer this question for youth ages 13-17. Schoenbeck et al.'s work on youth online harm resolution in platform-driven moderation demonstrated how top-down structures were incompatible with this age group's preferences for restorative justice~\cite{schoenebeck2021youth}. Juxtaposing these results with Seering's work on platform-driven versus community-driven moderation~\cite{seering2020reconsidering} reveals an opportunity for the latter to address this age group's need for highly contextual moderation decisions. In this way, the present work is aligned with Tekinbaş et al.'s studies of Minecraft communities, which researched how community-driven moderation influenced children's social experiences, but the participants of this server were significantly younger (ages 8-13) and the server was moderated by research study coordinators~\cite{tekinbas2021designing}. Self-determination theory and adolescent developmental psychology suggest that high school-aged individuals seek greater autonomy and independence and that adult-led interventions are less effective than peer-led ones~\cite{ryan2020intrinsic, deci2000and}. Additionally, work by Wiesniewski et al. and Ghosh et al.'s work have shown the benefits of participatory, resilience, and autonomy-based approaches to teen safety ~\cite{wisniewski2015resilience, wisniewski2017parental,ghosh2018safety, wisniewski2015preventative}. In this work, we hypothesize that empowering teens to moderate their own online communities can be promising for enabling healthier youth social media experiences.

\section{Methods}
In this study, we sought to understand what makes healthy and engaging teen-led online communities successful today. We emphasize that our study is not intended to represent the experiences of all teens and teen moderators. Rather, our goal was to examine the potential benefits, factors, challenges, and opportunities that result from supporting teens to lead these online communities. We chose Discord as our platform of interest because of its popularity among youth~\cite{gottfriend2023teens}, recent growth from the pandemic~\cite{browning2021}, and highly customizable moderation tools~\cite{discordmoderationtools}.

\subsection{Site of Study}
Our site of study was the \textit{Moderator Mentorship Community} (MMC), a Discord server created by platform staff designed to match teen-aged (13 to 17 years) moderators with young adult mentors and provide an open discussion space for youth community leaders. It was a small and highly selective program: interested teens had to pass the Discord Moderator Academy (DMA) Exam\cite{DMAexam} and apply for membership. Acceptance criteria for the MMC were not made publicly available, but participants relayed that the application included questions about demographic information, moderation philosophies, bots and tools used, and past moderation experience.

\subsection{Recruitment}
To be eligible for this study, interested teen participants from the MMC had to moderate at least one large online community with over 1,000 members on Discord. The minimum community size requirement was based on similar practices in past moderator studies~\cite{seering2019moderator}, which were designed to ensure that participants had enough relevant experience with regard to our research questions. Discord's community size distribution falls on a ``long tail'' curve~\cite{guo2009analyzing}, and past literature has shown that critical mass is important to the success and longevity of an online community~\cite{raban2010critical}. The two MMC mentors were recruited via referral from Discord staff collaborators.

\subsection{Participants}
Of our 13 participants, 11 were teenagers from the MMC, and two were vetted young adult mentors who volunteered in the MMC. Our participant pool overrepresented the Western and Global North, as all participants were located in Europe or the United States except for one teen from the United Arab Emirates. Our sample was also majority cisgender and male; nine participants identified as men, two as women, and two as nonbinary. Most participants started moderating around the age of 13, and a few began at age 12 (This was a violation of Discord's Terms of Service, which we discuss later.)

Participants moderated a wide variety of communities, and a majority were leaders in multiple servers simultaneously. Although participants were recruited via Discord, most disclosed that they also moderated communities on other platforms such as Twitch, Reddit, Minecraft, and YouTube. Most participants' communities were centered around topics related to gaming or entertainment, likely because Discord's history is rooted in gamer culture.\footnote{https://www.thedrum.com/opinion/2022/05/17/how-discord-became-the-center-the-universe-gamers} Other topics included self-help, education, livestreamers, professional esports players, musicians, and graphic design. Notably, 75\% of our participants moderated for at least one ``official'' community (i.e., endorsed by a commercial business) such as game development studios, music agencies, and esports organizations. We speculate that the MMC selection criteria might have favored Discord Partner servers, which was another application-based program that rewarded select comunities with unique branding, recognition, and perks for being exemplary communities according to platform standards~\cite{discordpartner}. 

Due to time constraints, participants were asked to focus their responses on the first community they started moderating in, as well as the community they currently participate in the most. Thus, the following table is not an exhaustive list of participants' experiences and only represents the Discord servers from our interview data.

\begin{table}
\resizebox{\textwidth}{!}{%
\begin{tabular}{|
>{\columncolor[HTML]{C0C0C0}}l |l|l|l|l|l|l|}
\hline
             & \cellcolor[HTML]{C0C0C0}\textbf{Starting Age} & \cellcolor[HTML]{C0C0C0}\textbf{Current Age} & \cellcolor[HTML]{C0C0C0}\textbf{Gender} & \cellcolor[HTML]{C0C0C0}\textbf{Location} & \cellcolor[HTML]{C0C0C0}\textbf{Community Topic(s)} & \cellcolor[HTML]{C0C0C0}\textbf{Community Size(s)} \\ \hline
\textbf{P1}& 13                                            & 15                                           & M                                       & UAE                                       & Musicians (Official)                                & 1k+                                                \\ \hline
\textbf{P2}  & 12                                            & 16                                           & M                                       & USA                                       & Self-Help                                           & 2k, 57k                                            \\ \hline
\textbf{P3}  & 13                                            & 15                                           & M                                       & USA                                       & Minecraft                                           & 5k, 200k                                           \\ \hline
\textbf{P4}  & 14                                            & 16                                           & M                                       & USA                                       & Games (Official)                                    & 85k                                                \\ \hline
\textbf{P5}  & \textless{}13                                 & 15                                           & F                                       & UK                                        & Esports (Official)                                  & 12k                                                \\ \hline
\textbf{P6}  & 13                                            & 17                                           & NB                                      & UK                                        & Streamers (Official)                                & 7k, 17k                                            \\ \hline
\textbf{P7}& 15                                            & 17                                           & M                                       & USA                                       & Games (Official)                                    & 220k                                               \\ \hline
\textbf{P8}& 16                                            & 16                                           & M                                       & USA                                       & Reddit Moderation, Cats                             & 3k                                                 \\ \hline
\textbf{P9}& 13--17                                        & 13--17                                       & NB                                      & Europe                                    & Games (Official), Streamers (Official)              & 4k, 18k, 200k                                      \\ \hline
\textbf{P10}& 13                                            & 16                                           & M                                       & UK                                        & Musicians (Official)                                & 3k, 6k, 8k, 67k                                     \\ \hline
\textbf{P11}& 13                                            & 16--17                                       & M                                       & UK                                        & Games, Streamers (Official)                         & 174k                                               \\ \hline
\textbf{P12 (MMC Mentor)}& 18                                            & 24& M                                       & UK                                        & Games, Streamers (Official)                         & 1k+                                                \\ \hline
\textbf{P13 (MMC Mentor)}& 13                                            & 20& F                                       & France                                    & Musicians (Official)                                & 100k                                               \\ \hline
\end{tabular}%
}
\caption{Information about participants' demographic background such as their age when they first started moderating, their current age, gender, and location, as well as the types and sizes of some communities they moderate. Interviewees answered with varying levels of specificity. Communities marked as ``Official'' are ones that were affiliated with a third-party commercial business}
\label{tab:my-table}
\end{table}

\subsection{Interviews}
Interviews were semi-structured and held via Discord. Ten participants were interviewed through a private voice call, and two opted for a written format through direct messages. We offered the latter to be more accessible in our methods since some users may not be comfortable with synchronous voice calls~\cite{mack2022anticipate}. Throughout the recruitment and interview process, we took care to recognize our position as adult researchers speaking with youth online and made this boundary clear in our communications. 

For the first 9 interviews, our IRB protocol required participants to obtain and send proof of their parents' consent via email. However, this deterred a significant number of prospective interviewees who did not want to disclose their moderation activities to their parents. A few said that this barrier would likely produce a biased and privileged sample of teenagers who had supportive and educated parents; another was concerned that their parents' email addresses would expose their last name. In response, we worked with our IRB to establish a procedure that allowed 13 to 17 year old moderators to participate without parental consent. The new protocol was approved given that the study presented minimal risk of harm and did not collect personally identifiable information. Participants were instead provided assent forms with age-appropriate language about the study procedures, informed that they could stop participating at any time, and compensated with a \$15 USD (equivalent in local currency) gift card for their participation. The first author conducted 10 of the interviews, and the last author conducted 3 of them. Some participants were asked clarifying questions via direct messages on Discord after the initial interview had ended. A final version of the interview protocol and codebook is shown in the Appendix.\\
\\

\subsection{Analysis}
We drew from constructivist grounded theory to structure our recruitment, interview, and analysis methods~\cite{charmaz2017constructivist, mills2006grounded}. In conducting theoretical sampling, our data analysis and collection often happened simultaneously~\cite{draucker2007sampling, breckenridge2009demystifying}. We did not recruit to obtain a representative sample of all teen moderators, but rather to reach data saturation for our research questions. Throughout the interview period, we met every week to discuss notes, extract themes, and iterate on our protocol. Interview questions were also influenced by current events and new legislation related to youth online safety at the time such as the U.S. state of Utah's attempt to enforce evening curfews for youth on social media\footnote{https://socialmedia.utah.gov/wp-content/uploads/2023/10/Social-Media-Regulation-Act-Proposed-Rule.pdf} and California's Age Appropriate Design Code Act.\footnote{https://californiaaadc.com/}

We drew upon Braun and Clarke’s methods of reflexive thematic analysis to review our data~\cite{braun2006using,braun2019reflecting}. The first author read and annotated the interview transcripts to get familiarized with the data and wrote memos throughout the process~\cite{thornberg2014grounded}. Then, the authors discussed these annotations and memos to jointly generate an initial set of codes~\cite{saldana2009coding}. This codebook was refined through an iterative coding process over several months. The first author coded roughly 20\% of interviews at a time, with each iteration discussed with co-authors to merge, remove, or add new codes. The themes in our interview protocol and codebook were not purely inductive: namely, questions and codes about child safety, parents, metaphors in moderation~\cite{seering2022metaphors}, and monetary payment were added deductively based on existing research and relevant news since participants were unlikely to raise these topics unprompted. After six months, the authors reached a consensus on a final codebook, and the first author coded the full set of interviews to produce the following findings. A final version of the codebook is shown in the Appendix.\\

\section{Findings}
Our findings are organized according to our research questions. We first focus on the experiences of teen moderators and interactions between members within their communities. Participants reported gaining many \textbf{skills and benefits} from their online community activities both as members and moderators. Many described developing valuable social, emotional, and practical skills, as well as increased exposure to diverse cultures, political philosophies, and career opportunities. Next, we present quotes about the \textbf{leadership values and moderation philosophies} that were commonly favored by participants and their peers in their communities. Teen moderators' emphasis on concepts like authenticity and respect suggested that their communities would not be the same had they been led by outsiders. Together, these two sections highlight that the peer-led nature of these online spaces is indeed central to their success and appeal, since they facilitated participants' lived experiences and genuine investment in their communities.

In the second half of our findings, we shift our focus to situate the many unique stakeholders involved with our teen moderators' communities. As noted, several MMC participants moderated for \textbf{third-party businesses }such as game developers, influencers, and artists. These partnerships with official organizations can be highly beneficial and appealing for young users, but they can also introduce difficult power dynamics.  We then explore\textbf{ the role of the MMC}, which consisted of peer networks, dedicated mentors, and platform resources to keep teen moderators and their communities safe. Finally, we detail the many \textbf{risks and challenges} that participants still faced despite having these external support structures such as work-life balance, harmful content, harassment, and lack of understanding from their offline community. These three sections illustrate the role of peers, stakeholders, and platforms that make these youth-led online communities possible, as well as several opportunities for better support.

\subsection{Experiential learning: Social, emotional, and cultural development}
\subsubsection{Interpersonal and communication skills}
Online community moderators on Discord are responsible for enforcing rules, developing culture, and growing the community~\cite{seering2019moderator}. In any social experience, conflict is bound to happen, and in participants' Discord communities, moderators were usually the ones responsible for resolving them. Interviewees expressed that they developed strong social, emotional, and communication skills through these experiences over time:
\begin{quote}
\textit{``Moderation has helped me understand a lot about how to deal with people, especially with regards to managing conflicts, resolving them, working with different types of people [from around the world] ...  It has basically helped me a lot in terms of the sort of soft skills that you need to run a community smoothly.' ---P4}
\end{quote}
This growth was not limited to the moderators leading the communities. Participants explained that when addressing conflicts, they often took an educational and reformative approach, taking care to explain infractions to offenders and hoping to impart ripple effects on the rest of the community. These skills even transferred to their lives and friends offline:
\begin{quote}
\textit{``I think the way that I've handled things through Discord, such as arguments, also kind of benefit me in life. So, of course, I'm going to get an argument with friends. That's a healthy relationship. And I kind of use the skills through Discord in real life to de-escalate situations.'' ---P11}
\end{quote}
\subsubsection{Exposure to global and diverse cultures}
Discord communities can encompass members from all over the world, and participants frequently encountered perspectives and backgrounds that diverged from their own~\cite{kahne2011diverse}. Being exposed to global communities at an early age helped them understand and respect for other cultures. Multiple participants recounted stories of instilling this belief in others in their community. P1, for example, was located in the UAE and had navigated these cultural clashes many times:
\begin{quote}
\textit{``Sometimes someone from the Western side of the world would say something, and then someone else would think it’s weird or disrespectful, and same the goes for the other way around [...] So I try to help them and share that this person didn’t mean anything bad, and they  were saying[it] with good intentions.'' ---P1}
\end{quote}
\subsubsection{Developing philosophies of justice and politics}
Teen moderators develop philosophies about justice and politics, even if the community's topic is not expressly political in nature. For example, P3 explained their approach when resolving conflicts in their Minecraft community. They tried to always give each side of the situation an equal, unbiased, and private channel to explain themselves. They were also aware of how individual offenses are not always isolated incidents, but could actually be symptoms of deeper and systemic problems with the community:
\begin{quote}
\textit{``[After resolving a conflict], I talk to [the offender] to make sure that they're feeling welcome in the server and [make sure] it's not because \textit{of the server that they're acting out.'' ---P3}}
 \end{quote}
Although participants did not name it as such, many were practicing similar methods of procedural justice when engaging in conflict resolution. The theory of \textit{procedural justice} asserts that people's perception of a justice system is strongly impacted not just by its results, but also by the quality of their individual experiences with it~\cite{lind1988social,meares2017policing}. It emphasizes the importance of principles like due process and right to trial. It also presents at the organizational level; according to Meares et al., a community's trust in an enforcing power depends on how it presents itself in the public eye~\cite{meares2017path}. P3 demonstrated an intuition for this concept when they explained how an entire server's moderation team loses credibility when individual moderators disagree on an issue publicly:
\begin{quote}
\textit{``I think [picking sides in a conflict] can be a very big issue that can lead to moderation disagreements and [result in community members] having these bad relationships with moderators ... It can really divide a moderation team if you do.'' ---P3}
\end{quote}
\subsection{That's capital: Technical skills and career opportunities}
\subsubsection{Self-motivated technical learning }
Past scholars have studied how youth participatory culture promotes informal learning, increases digital media literacy, and strengthens civic agency~\cite{ito2013hanging}. Our data supports this literature, but as moderators, the impact and influence participants had on the community seemed to motivate them even further. P12, a mentor, reflected on how they felt when they first started moderating:
\begin{quote}
\textit{``You still feel like you're in this teenager mindset and---it's really weird, I can't explain it---so you're doing this fun hobby stuff, but there are different roles and positions and places and responsibilities that came with that. I think the appeal for it as a young person is [that] what you do matters. You can feel [your] impact on how the community operates, and the vibe and the personality that it has.'' ---P12 (MMC Mentor)}
\end{quote}
This motivated many participants to pursue various technical skills to improve their communities. P6 and P3 learned to code at an early age so that they could build custom Discord bots, and both now teach younger developers in software engineering mentorship communities through paid and unpaid opportunities. 
\begin{quote}
\textit{``About a year ago I started to learn Rust and Golang, which are very useful. Golang used to make microservices, which bots written JavaScript can then easily hook into ... I've learned those also online, the internet's a great tool for learning ... I now tutor people in programming JavaScript for a chain company for people from the ages 5-14 to learn to code ... and by teaching, that’s further solidified my knowledge.'' ---P6}
\end{quote}
P2 similarly developed an interest in image editing and graphic design through their Discord moderation activities, and the latter eventually established a community of practice where experienced members teach newer ones skills like creating server banners, themes, and emotes. 

\subsubsection{Discovering career paths and paid opportunities }
Moderation sometimes opened doors to direct career and job opportunities. P5 relayed that some of the older moderators in their team were hired as full-time employees for the professional esports organization they volunteered for. P10 aspired to pursue a career in community management and was actively building experience in smaller communities to make their resume competitive for larger paid opportunities in the future.

A few participants had already earned money for relevant skills and expertise. P6 was well-known in many developer communities for their ability to make highly customized Discord bots and sometimes charged comission fees for their work. P10 had been hired for temporary moderator support for communities for approximately £100 per month before, and they had also provided growth strategy consulting for various communities and charged around £40 per session. They explained what these consultings sessions entailed:
\begin{quote}
\textit{``We sit down for a few hours and they just walk me through their server’s problem and I give them advice like what kinds of [moderator] roles they have/should have ... I'll also look at their community size and type of engagement, look at what they want to change. It depends a lot on the community. Some just want to grow more members. But some might be looking to make their communities healthier. Sometimes they are even trying to shut down the community or demolish it and need advice on it, things like that.'' ---P10}
\end{quote}

\subsubsection{Getting extracurricular credit and volunteer hours}
Participants also found ways to weave their moderation activities into their school extracurriculars. For example, P6 was making their EPQ (Extended Project Qualification)\footnote{EPQs are a type of high school extended project and qualification exam in some regions of the UK.} about volunteer online community moderation research, theory, and practice. 

P2 was a leader in a unique community that enabled members to earn volunteer hours online. The owner of the server was a high schooler who started the Discord community as an online peer tutoring space during the peak of the COVID pandemic. At the time, in-person volunteer opportunities were largely cancelled, so the founder took the initiative to register the server as a 501(c)(3) non-profit. This process allowed high school students to officially earn service hours for tutoring peers online, which is a requirement for many public institutions in the US. Participants could also include these service hours in their job and college applications to boost their specs. 
\begin{quote}
\textit{``This is probably, well actually it is the biggest one that I'm in right now and it has a total of around 57,000 members. And it's basically a server for students, middle school, high school, or college to just go there, meet other students, engage in meaningful conversations, and also get tutoring, homework help, notes, learn about program, there's so many different things that we have. And there's also since it's a student-run nonprofit, it's actual organization.'' ---P2}
\end{quote}
Another way to earn hours with this organization was to volunteer as, of course, a Discord moderator. Teens performed various leadership tasks like holding online debates, facilitating TED Talk discussions, and other virtual peer learning events. In addition to giving members opportunity to earn formally recognized credits, the server offered teens a preview of professional work and business culture. Many interviewees responses about the moderator hiring and application process revealed similar organizational practices like interviewing, networking, and resume crafting~\cite{wanberg2020job, ito2009living, marwick2011tweet}. P2 further elaborated that this was a great community for students precisely \textit{because} it was ran by students:
\begin{quote}
\textit{ ``[The moderators] offer staff positions to students and since it's also ran by students, it's a great opportunity for students to gain experience in things that they're interested in ... I've been a moderator for a few weeks now, and I used to be a community manager last month. I've been switching around [different roles]. Today, I just got promoted to a new position. I'm still gaining experience, learning new things from it, and it really is a very professional environment, at least the staff team. It really gives us students a feel of what it's like to work at a big company and get these positions, network, all of that.'' ---P2}
\end{quote}
\subsection{By youth, for youth: Peer leadership promotes participation}
In addition to understanding the benefits that teen moderators gain in teen-led spaces, we are also interested in examining what factors and values might motivate youth to voluntarily participate in these peer-led online communities. In this section, we present quotes that allude to why being designed by youth, for youth makes these Discord servers appealing and fun for users in this age group.

\subsubsection{Authentic cultures of respectful, moderated engagement}
A key common trait among participants' communities was that they were about topics that were genuinely fun, interesting, or enjoyable to teens. Participation was not forced; community members joined because they wanted to. Sometimes this was a result of having existing fanbases from elsewhere, such as ``sibling''  Discord servers of Reddit, Twitch, YouTube, and Minecraft communities~\cite{seering2019moderator, alcala2023altspace}. Others joined because they liked the central figure or personality, such as P5 who moderated a famous esports player and streamer's Discord server. Of course, gaming communities were overrepresented on Discord, and we remind readers about the MMC's potential bias for Partnered servers.  However, we contend that members' authentic and genuine engagement was key to these communities' success and appeal. This is an inherent and somewhat obvious characteristic of community-based platforms, but its effects on the community culture and ``vibe'' cannot be overstated. For example, P2 explained that even though their peer tutoring server was school-related, they believed that the Discord server facilitated more vulnerable and honest conversations than anything they could find on campus:
\begin{quote}
\textit{``I think that through Discord, there can be meaningful conversations on relationships, teaching people, connections, that aren’t found in schools ... [Our Discord community is] providing that space to be able to [talk to others] in a way that's like, a healthy space. If anything goes wrong, there’s actually something [that's going to be] done about it instead of just like being like, `It happens.' ...[The fact] that [disagreements] are communicated about, and the respect, those are key factors to our group.'' ---P2}
\end{quote}
\subsubsection{Emergent leadership and earned legitimacy}
Most participants explained that the best communities are the ones where the moderators were actually part of the community and had a solid connection with members. This aligns with Seering's social identity theory applied to online communities; in most groups, the leaders that emerge are prototypical of the general community members~\cite{Seering2018applications}. For example, P4 elaborated that moderators who actually play the community's game are more effective and respected:
\begin{quote}
\textit{``I think that people are most effectively able to moderate if they actually care about the community ... [It's also very important] to be a role model to the community. When you're more invested, the community is going to respect you more and listen to you more than someone who has absolutely no idea how the game works. They're obviously not going to listen to what they have to say.''---P4}
\end{quote}

Related to these themes of respect, participants asserted that being a moderator does not actually mean anything on its own. To have power, one must earn the trust of others in the community. These models of governance strongly imply that the self-driven and community-based structures of are indeed consistent with middle adolescent needs for respect and autonomy, further supporting that peer-led Discord communities might be more appealing to youth.

\subsubsection{Distaste for top-down authority and impersonal moderation}
Conversely, participants had negative opinions about opposite styles of leadership. When asked about which metaphors of moderators resonated with them among choices like  ``dictators'', ``police'', ``mediators'', or ``janitors.'' ~\cite{seering2022metaphors}, most said they would never want to be seen as any of the descriptors in the authoritarian categories. Participatnts strongly disfavored top-down approaches to moderation, and P1 modeled their leadership based on negative examples of authority they had observed in past servers. 
\begin{quote}
\textit{``A lot of times when I join Discord servers, the staff team would [be] so annoying ... and they wouldn’t just respect you as a member, they would just abuse their role and abuse their power... So in my server, I decided to have a staff team that reflects the members... they would moderate the server and make the members feel at home so everyone would talk together; they would chat and make friends, no problems at all.'' ---P1}
\end{quote}
These findings suggest that communities run by outsiders may indeed not be appealing for teens. Many also disfavored styles of moderation that lacked a personal or human touch:
\begin{quote}
\textit{``It's a lot better [to have someone invested in the community] than getting someone with no ties to the community and they're paid to just enforce the rules because I think at that point, just get a bot and get AutoMod or whatever. You don't need to have a person to do that.'' ---P4}
\end{quote}
According to P12, good moderators should instead know when to ``bend the rules'', and when a moderator has a ``strict or black-and-white style of moderation'', it signaled that they are not actually invested in the community. Other bureaucratic practices like punishment quotas were also disfavored. In the next section, we present examples of how these styles of moderation were often enforced by adult stakeholders when they entered previously teen-led spaces. 

\subsection{Interactions with adult stakeholders from third-party businesses}
Like most other social media experiences, Discord communities and moderation teams are not segregated by age group. This means that online communities can be any mix of youth and adult members, and we learned that many of the participants in our study regularly interacted with adult stakeholders who represented third-party businesses. These official collaborations had clear benefits for all parties involved, but they could also introduce complex power dynamics due to differences in values, goals, and age.

\subsubsection{Benefits of partnering with official organizations}
Many of our participants were moderators for fan communities that had official collaborations with relevant third-party businesses. This meant that they were endorsed, owned, or managed by businesses such as game studios, internet personalities, or professional musicians. Having the support of official organizations could immediately give a community more popularity and clout by enabling opportunities like in-game rewards, interactions with an influencer, or access to exclusive information:
\begin{quote}
\textit{``It's a lot easier for your community to make events regarding a game when the developers are on your side ... In the server I currently moderate, the owners do have contact with the developers and we're actually able to hand out monthly gift codes to users that let them claim in-game rewards. This attracts a lot more people to our server because there would be a lot less people here if it weren't for the free stuff.'' ---P4}
\end{quote}

These partnerships sometimes also resulted in occasional small gifts for volunteer teen moderators in the form of tokens such as in-game currency, items, or Discord Nitro boosts. Official collabrations were invaluable assets for businesses too, since teen moderators who understood their target demographic could help them build a more authentic brand, voice, and connection with young users. This was vital to modern online marketing strategy, P11 explained:
\begin{quote}
\textit{``Lately it's all about how smaller conversations are happening and how smaller Twitch streamers are more relevant than the big massive corporate streamers that are out here shoving their partnerships down your throat and their advertisements for days and sponsorships everywhere.'' ---P11}
\end{quote}
\subsubsection{Navigating power dynamics and differentials}
Though these partnerships could be beneficial, when adult stakeholders from external businesses entered previously volunteer-led youth communities, it often also introduced complex power dynamics. Business representatives often lacked context and did not operate on the same values as teen moderators. P10 described that it was painfully obvious when outsiders were being fake and did not understand the culture.

In P4's server, these problems were less obvious and instead resulted from the top-down, hierarchical leadership structure. Game developers in their community typically only interacted with the admins, who were moderators with higher permissions and seniority. This lack of communication was frustrating since sharing information in advance could have saved the larger moderation team significant time, stress, and effort:

\begin{quote}
\textit{``Most interactions are just with the admins and the developers. I personally don't really like that because it feels like they're always keeping information from [the rest of the moderation team], as if some elitism between them and the developers ... The first time we had a major game update and we weren't really made aware of it, the communication[to the general community] didn't go out super well.'' ---P4}
\end{quote}
Teen moderators in these tense relationships did not feel empowered to speak up about unfair treatment because they recognized their position as unpaid volunteers. When asked whether they would feel more appreciated if they were rewarded with in-game gifts, for example, P4 explained:
\begin{quote}
\textit{``I feel like it's definitely not my place to ask the developers for stuff in the game just because I moderate their server. I personally don't care much because I see moderation as more of a hobby than a job and would rather not be paid for it. At the same time, I feel like it's a bit unfair when we're doing the most work and not getting as much recognition.'' ---P4}
\end{quote}
\subsubsection{Age disclosure and adult stakeholder bias}
Another unique challenge of working with third-party businesses was that external stakeholders often treated moderators and users differently based on their age. P12, an MMC mentor, observed that this was apparent even in Discord's interactions with teens, and that businesses understandably want to be extra careful with policies regarding children:
\begin{quote}
\textit{``Discord just [doesn't] want to get anywhere near the child labels. So they'll avoid it, which is really annoying and frustrating... There was a little bit of paranoia about what can be done legally...so I was just thinking, `Okay, Discord really takes this seriously. They really don't want to mess up anywhere here.' ...There was a lot more red tape.'' ---P12 (MMC Mentor)}
\end{quote}
It was not always clear if this treatment was simply out of legal precaution or if it was the result of prejudice and ageism. P10 recounted how they were initially approached by a music agency's team to become a moderator because of their reputation as a standout member, but upon learning that P10 was 16 years old, the employee almost revoked the offer.
\begin{quote}
\textit{``A lot of people actually will approach me first and offer me a mod position but then change their mind once they learn my age. And I’m like, you can’t just do that… But some people will judge you based on your age no matter what.'' ---P10}
\end{quote}
This bias was not as evident in peer-only interactions between youth. In fact, when asked how old their moderator teammates were, P12 found it amusing that ``[outsiders tend to] forget that there isn't a name badge [next to our Discord] IDs saying how old someone is.'' Participants explained that they usually assumed their fellow community members and moderators were near their age and treated them as equals (unless otherwise indicated, such as by employee role titles). For example, when asked about the qualities they look for in potential moderator teammates, P10 explained that an applicant's age was rarely ever relevant and that they assessed candidates by the explanations of their thought processes in hiring interviews. There were some exceptions like if a user had a strict minimum age policy for moderators\footnote{Some communities require moderators to be 18 or older due to legal concerns of affiliated businesses or the potential of encountering adult content, such as in fan art servers. None of our participants reported moderating for these types of communities.} or if an applicant was under 13 years old, which violates Discord's Terms of Service, in which case participants would usually report them to the Discord. Still, several participants confessed to becoming moderators at the age of 12 or younger in past communities by lying in their applications, interviewing well, or succeeding as ``trial'' moderators which are like temporary internship roles. P2 admitted that they should not have been allowed to join in hindsight:
\begin{quote}
\textit{``You're not supposed to be on Discord until you're 13, but I didn't know that at that time ... I'm pretty sure I was still 12, but despite me being maybe a few months before the age requirement, they [community moderators] decided to bring me on board anyways, and I really felt grateful for that. But looking at it now, they shouldn't have done that.'' --P2}
\end{quote}
\subsection{The role of the MMC: Peers, mentors, staff, and platform support}
Teen moderators in our study existed in a complex ecosystem and relied on many networks of support to keep themselves and their communities safe. Like other moderators, our participants worked in teams and frequently interacted with other moderators in their communities to perform routine tasks and make informed decisions~\cite{seering2019moderator}. Many also referred to documentation written by senior teammates~\cite{seering2019moderator} and consulting resources like the Discord Safety Library. However, the most valuable system according to our participants was the MMC.

\subsubsection{Meaningful peer discussions and exchanges}
Participants emphasized that much of the MMC's value came from the community members themselves. The peer discussions and interactions with teen moderators from other servers were thought-provoking and informative:
\begin{quote}
\textit{``We discuss moderation tactics, best practices for keeping ourselves healthy, keeping communities healthy, make ways to make our lives easier, ways to make sure that we're doing the best job possible ... It's really, really useful to have mentors and people that are way more experienced than me to guide me through different elements of Discord moderation that I've never delved into before.'' ---P3}
\end{quote}
P3 felt similarly, and they speculated that their interactions with others in the MMC were related to how rigorous the selection criteria were. They shared that according to rumors, over 800,000 people took the DMA exam, but only a few hundred teen moderators made it in to the MMC. Based on this information, they explained that everyone in the community must have studied the Discord Safety Library articles, had experience moderating a successful community, and---most importantly---were there because they wanted to learn and make connections with others. The peer interactions were so important that P4 noticed that the lack of newer MMC recruits was making the community stale:
\begin{quote}
\textit{``Initially, I found it to be very helpful as mostly it was because of the other people in there ... there was just a really good exchange of information that I'm sure helped all of us out on our journey. At this point however, since it's been many months since new people have been added, activity has really died down ... I didn't want to learn from just} looking \textit{at other moderators. I wanted to learn from talking and discussing with other moderator ... At this point, the server in general has become very inactive. There's not much we do anymore.'' ---P4}
\end{quote}
\subsubsection{The unique perspectives of near-peer mentors}
Like the name suggests, the MMC also included a formal mentorship program. When teen moderators joined, they were matched with near-peer mentors. These mentors were young adults from the DMD who volunteered for the role and were vetted by the MMC staff. As part of this study, we interviewed two MMC mentor participants about their perspectives on their role and the MMC program at large. P12 explained that it was a smart move on Discord's part to establish these peer and near-peer relationships, since it could be difficult for platforms to establish meaningful conversations with teens directly:
\begin{quote}
\textit{``[It's really important to] be able to put in the right resources and know what the right triggers are for starting healthy conversation and discussion, because it can feel a little bit alienating if you're not used to it. It can feel a bit like the, `Well, hello fellow kids,' kind of meme if you are not sure what to talk about. You need to build a rapport with them.'' --P12 (MMC Mentor)}
\end{quote}
We also asked them to elaborate on differences they observed between the MMC and DMD, since they were in both. According to P13, our other mentor participant, the MMC seemed to facilitate a more respectful and educational environment. She hypothesized:
\begin{quote}
\textit{``I think the MMC is a better place [for the teen mentees] since there is a hierarchy in the server. When you are a mentor, they tend to respect you as a teacher. I think because of that classroom whole ambience, they are more respectful towards each other, too ... Also, as mentors, we try to initiate both educational and casual conversations, which really helps to make it a nice place for them.'' ---P13 (MMC Mentor)}
\end{quote}
P13 further explained that this was very different from the culture of the DMD, where many adult moderators behaved as though they were superior and more experienced than others. They were glad that the MMC existed as a separate place since this negative environment could be very discouraging and unhealthy for young learners.

\subsubsection{Access to platform staff and opportunities}
Another key benefit of the MMC participants cited was that they had more access to Discord staff and support. It was a rare opportunity to have that connection with the platform at all, and participants considered this to be a huge privilege. Being a part of the MMC, one participant observed, opened doors to participant in research like this one.
\begin{quote}
\textit{``It’s a real privilege to have access to it. I’ve been accepted into a few student councils, but the MMC has a lot of studies [like this one] being done. And I really think it is a privilege to be here with all the research,  it’s played a huge role in all my skills and experience. And I really can’t wait for it to grow more. I know Discord’s Safety team has a lot planned.' ---P2'}
\end{quote}
MMC membership also gave participants a visible badge on their Discord that gets displayed on their profiles to all users. This type of reward has been noted and suggested in past studies as a form of appreciation~\cite{seering2022pride}, and according to one participant, it did more than just serve as a token of recognition. The MMC reputation was so positive that simply having the badge and including it in their application gave them an instant boost when they applied to moderation teams in other communities. 

Almost all participants acknowledged that programs like MMC servers were highly exclusive and hoped similar resources would become more widely available, especially since publicly availables ones were usually not very useful. For example, the public Discord Town Hall server was described as useless, and feedback never seemed to go anywhere:
\begin{quote}
\textit{``The Discord Admins Server is the most direct way [to provide feedback]... While they do respond, it’s usually just like brushed off or ignored. I just gave up.'' ---P10}
\end{quote}
\subsection{Challenges with moderation that teen participants still face}
Thanks to the MMC and their many other support systems, our participants are likely among the most well-supported teen moderators on Discord. Despite their background, however, there were still several challenges that participants experienced. Our interviews confirm that teen moderators do face the risks documented in past studies about community-based voluntary moderation such as struggles with work-life balance and exposure to harmful content, and the repercussions are arguably worse for teen volunteers given their demographic. 

\subsubsection{Struggling with work-life balance and burnout}
The most common challenge participants discussed was ``work-life balance'' which referred to difficulty managing one's time and energy as a volunteer. P1 recounted how they had once let moderation take over their life, which caused social, academic, and mental health distress:
\begin{quote}
\textit{``I got really addicted to talking on Discord, to moderating and running servers. It just got to a really bad point where my family doesn’t see me, my grades have dropped, and there are many negative effects from Discord ... My friends, my family, all of them have noticed.'' --- P1}
\end{quote}
This was especially salient for teens who felt that being a moderator was not a compartmentalizable activity. To many participants, being a good moderator meant being actively and authentically engaged in the community, and taking extended breaks or treating it as a part-time job on the weekend was not feasible. A few described that they had found a healthy balance between moderation and their other activities, but it required a strong sense of self-awareness and discipline. 
\begin{quote}
\textit{``I have a set amount of time that I've spent doing it, and I'll avoid doing more than that because that does then end up increasing stress and have a negative impact on mental wellbeing, such as my sleep schedule ... I've been treating it less as something that I have to do, and recognizing that it is actually a voluntary position. I can take a step back, I can assess things. If I feel like I'm taking on too much work, then I will reduce that workload.'' ---P6}
\end{quote}

While on the topic of time-management, many elaborated that they hoped to continue moderating in the long-term but knew that they would eventually have to step away as they entered new stages in life as they got older such as going off to college or finding a full-time job. Participants noted that some senior moderators in their teams found balance by taking on higher-level admin roles that required less time and energy, and hoped to at least be able to do the same.

Relatedly, we asked participants for their thoughts on the idea of getting paid by Discord or third-party businesses for moderating. While some thought it would be nice to get paid, others conjectured that it could result in undesirable practices like logging hours or fulfilling ban quotas~\cite{seering2019moderator}, and a few were concerned that people might start moderating for the money instead of out of a genuine interest in the community. 
\begin{quote}
\textit{``I think it could lead to an increased sort of... I guess an increased outside factor. I can't think of the word for that, but an outside factor that leads to people trying to fulfill quotas, or specific amounts of hours working for a server moderating its chat. And I think that can be dangerous, because it leads to people being punished for things that they haven't done, just so that the moderator can get a quota and be paid.'' ---P3}
\end{quote}
\subsubsection{Increased risk of encountering harmful content}
Harmful content such as hate speech and explicit imagery was the second most commonly discussed challenge, even though stakeholders would likely consider it the highest priority. The most concerning instance we learned of was P5's experience where they had once accidentally opened a disturbing photo while moderating. This participant was also one of the few participants who expressed not wanting to continue moderating in the long-term: 
\begin{quote}
\textit{``There was[a moment when] we got looped with gore stuff. That wasn't nice because you click on a video expecting it to be something related to the server, but it's something completely disturbing. Stuff you don't want see. It was pretty gory.'' ---P5}
\end{quote}
P12, an MMC mentor who had began moderating at the age of 18, expressed concern about the emotional effects on younger users who encounter these types of harms. They recalled a moment when a younger mentor's reaction to a piece of content was unusually mild in comparison to their own:
\begin{quote}
\textit{``I remember seeing some shocking content before and I would be speaking to one of my mentor peers and just talking to them about how upsetting it was, and they were far younger than me; about three, four years younger than me. And that was just kind of a moment of, `This person's so much more used to this than I am. And that's confusing. Is it that younger people today are more experienced with these situations and they're able to adapt faster than I am?' ... It's different for me because when I was a teenager, I just went outside, and I hung around on my bicycle with some friends and that was it.'' ---P12 (MMC Mentor)}
\end{quote}
Another safety issue frequently at the top of stakeholders' concerns is child safety and child sexual abuse material (CSAM). Despite these being Discord's one of top concerns for young users~\cite{discordTransparency}, the topic was never brought up by participants organically. When we asked interviewees for their thoughts on child safety, they assumed that the term referred only to users 12 years old or younger, even though platforms consider all users under 18 as children: 
\begin{quote}
\textit{``Do you mean like child predators or like really older people being manipulative? ... Um, all right ... If you go into the right communities with the right moderators and everything, then I think you'll be safer as long as you kind of know what you're doing ... But no, it doesn't happen very often, at least not from what I've seen ... Child safety is important, but I think that there's always going to be [people who find] ways to manipulate children. So I think it's just always best to like keep kids [under 12] away from the internet as a whole.'' ---P10}
\end{quote} 
The participant stressed that having strong peer networks and engaging only in well-moderated communities could help keep kids safe. Another agreed that it was an important issue but felt that attention given to it by the public was excessive:
\begin{quote}
\textit{``Child safety is an issue of course ... [but I think] it’s blown out of proportion. They way people talk about it makes it sounds like it happens every 5 minutes; it’s not like that ... When something like that does happen, I have friends I can go to and talk to about it. Also with Discord I think there are many ways to easily report someone.'' ---P11}
\end{quote}
These quotes are interesting and informative about our participants' experiences, but we note that our data does not represent all teens on Discord. One potential reason why interviewees did not frequently encounter child safety threats might be because these threats may be situated more often in private channels like direct messages (DMs)~\cite{freed2023understanding} whereas moderators are responsible for semi-public spaces.

The last type of harmful content in this category was threats of real-life or physical harm. P10 told a story about a stressful time when a community member would frequently DM him about their personal problems and suicidal ideations. P10, uncertain whether he should report the user or try to be a supportive listener, opted for the latter. The user eventually sought professional help, but P10 described that it was an emotionally exhausting situation.

Harmful content remains one of the top concerns about teen social media use, and our data suggests that teen moderators likely encounter it more frequently than the average teen on Discord. There are many platform systems and automated tools like bots in place that were helpful for dealing with these issues such as platform safety and reporting tools, meta-communities for cross-community bans, and safety news from various YouTube content creators. Of their options, participants most frequently described peers in their moderation teams and the MMC as the most valuable resources for support.

\subsubsection{Moderator harassment and negative stereotypes}
We discuss the topic of harassment separately from other forms of harmful content because participants were often harassed directly \textit{as a result} of their position as moderators. Like past studies have found, volunteer online community moderators are more likely to be targeted for harassment and insults as a result of their roles~\cite{seering2019moderator}. Many of our participants encountered harassment related to negative depictions of Discord moderators in mainstream media. These stereotypes were pervasive enough that when when we asked participants whether their friends in real-life knew about their moderation activities, several explained that they did not tell their peers offline because of these assumptions:
\begin{quote}
\textit{``There's this one GIF [that stereotypes] Discord moderators as a load of fat people. I don't know if you've ever seen it. I got sent that a couple times.'' ---P5}
\end{quote}
The targeted harassment was likely worse for moderators of marginalized backgrounds. P7 explained that they encountered many forms of harassment related to their gender identity and had developed their own way of coping with these experiences:
\begin{quote}
\textit{``There have been occasionally been users who have targeted me in particular, because I'm queer ... I've reported to Discord and banned them from my community. They occasionally come back to make an alt and target me again. But after a while, after two accounts or whatever, they just get bored and they move on, and you never see them again. Of course, I don't take it to heart; why should I take it personally if this person clearly just wants to troll communities? It doesn't impact me in any way. But I guess that exists, and it counts as a form of harassment. But I've dealt with that in the best way I think I can. So there's that.'' ---P7}
\end{quote}

\subsubsection{Lack of understanding from parents/guardians and offline community}
A final challenge we surfaced was a lack of understanding from participants' offline communities. Despite their extensive online networks, most participants felt that their parents, guardians, schools, and friends in real-life would not be supportive. A minority of participants said that their parents knew about their role as a Discord moderator activities, but most were neutral or did not really get what it meant. Only one participant had explicitly supportive parents and regularly discussed moderation at the dinner table. The majority of our participants explained that they did not disclose or discuss moderation with their parents/guardians. At best, participants simply did not want to have to explain; at worst, participants kept it a secret out of fear of judgment. P10 wished they could clarify these misunderstandings about social media and Discord with their parents:
\begin{quote}
\textit{``She knows I’m on Discord but she doesn’t understand what it is or what I do on it. To her, it’s just like, `You’re talking to strangers and it’s dangerous and you should never do it!' I think she knows that I mod for [a famous YouTuber] because I mentioned it to my little brother who also watches him, but I’ve never really had like a conversation where I sat down and explained it to her ... [I wish she knew] we live in a world where everything is on social media. Unlike [other apps], Discord is a lot more about connecting with other people and the community. It feels less dangerous than like Instagram and TikTok [which] can feel a lot more isolated or self-focused, but here it’s sort of like connecting with people, and finding a sense of belonging.'' ---P10}
\end{quote}

P4 similarly hoped offline stakeholders would someday be more open-minded. During the COVID pandemic lockdowns, they proposed incorporating Discord into their school's technology stack but recalled being rejected immediately: 
\begin{quote}
\textit{``I managed to secure a meeting with [my school's] administration, but the second I walked in, I knew that they were going to shoot the idea down because they didn't like new stuff. I guess they just didn't want to try something they hadn't already been using ... There should definitely be more awareness [among] older people like parents and school administrators about [how] new technologies can help better people's lives and build experience, like moderation, for example.'' ---P4}
\end{quote}
Support from adults offline could be more than just beneficial---it could even be life-saving. In one of P10's communities, a young Discord server owner was threatening self-harm in response to an incident of community backlash and bullying. The only way they were able to resolve the situation was by contacting the teen's parent who had always been very lightly involved in the server. Responses like these suggest that having the support and understanding of trusted adults could be critical to the safety of teen moderators and their communities. Thus, one of our key recommendations for stakeholders in the following discussion is to increase acknowledgment and transparency about this important population to better support, protect, and empower teen-led online communities. 

\section{Discussion}
In summary, our findings show that being designed both \textit{by} and \textit{for} teens is central to the benefits and appeal of communities like those run by our participants. The same level of skill development, identity capital, and genuine engagement that our participants facilitated would likely not be found in adult-curated youth online interventions. However, our interviews also demonstrate that this type of growth is difficult without support from various external stakeholders including peers, mentors, staff, businesses, and the MMC. This broader ecosystem was critical to the success and safety of our participants' communities. In our discussion, we discuss the implications of these findings in two ways. First, we provide a set of \textbf{recommendations for stakeholders} to better support, protect, and empower teen moderators and their communities. Second, we analyze \textbf{the role of Discord and community-based platform structures} in our participants' experiences to articulate why these offer promising healthy and engaging youth-led online social experiences. 

\subsection{Recommendations to better support teen moderators and their online communities}
\subsubsection{Platform acknowledgment, transparency, and research about teen moderators}
Today, there is little to no acknowledgment of teen moderators or the role they play on Discord. Formally recognizing teen moderators could make their activities safer, similar to how legalizing the labor of children and undocumented workers led to more protections~\cite{betcherman2005child,rivera19999undocumented}. These types of protections are urgently needed for teen moderators given the challenges participants described regarding harmful content, harassment, work-life balance, and third-party relations. The emotional effects that teen moderators may experience from encountering harmful content, harassment, and stress are serious given that adolescents may be more susceptible due to developmental sensitivities~\cite{orben2022windows}. Like others have urged regarding youth and social media at large, we contend that more transparency from a platform level is necessary~\cite{sala_social_2024}. To this end, more representative, large-scale studies with teen moderators are needed, especially about topics like child safety that our current data could not sufficiently address.

Increased recognition and transparency could also help mediate the power dynamics participants faced when working with third-party businesses. Although none of our participants experienced them, exploitative relationships between adult influencers and teen moderators have been covered by media in the past~\cite{jiang2023genshin}. Potential solutions might be to require background checks and training for adults representing third-party businesses in youth-related communities. Furthermore, the lack of equal treatment and respect due to teen moderators' age could motivate young users to lie about their age to bypass relevant policies. Instead of turning away from these potential risks, increased transparency and platform acknowledgment could improve the safety of teen moderators and their communities. 

\subsubsection{Recognizing and legitimizing teen moderators' skills and labor}
Our interviews confirmed that one of the most pressing issues from teen moderators' perspectives was work-life balance. Current guidelines provided by Discord focus on limiting the amount of time that moderators spend volunteering. Though well-intended, these recommendations to limit time spent moderating assumes that the activity is negative in excess. This has not been found to be uniformly or straightforwardly true, and in this study, we have shown that both positive \textit{and} negative outcomes are possible among heavily involved teen Discord users. Rather than instructing teen moderators to regulate themselves, stakeholders should either be providing more structured boundaries or ensuring that youth are getting value from volunteering.

Recommendations by past moderation research with adults have generally advocated for the latter in the form of social recognition or monetary compensation, but monetary compensation is a tenuous solution for minors, especially on a global platform, and risks barring teens from moderating completely. It also goes against many moderators' values regardless of age~\cite{seering2019moderator, seering2022pride}. Instead, there are many alternative ways stakeholders can ensure teen volunteer moderators are not left empty-handed. Badges, certifications, and other formalizations of skills and labor could be especially valuable for youth since, as in the case of P, they could give teens more credibility for future opportunities. Certifications like the DMA exam could also be brought back. If platforms worked with institutions to make them formally recognized like on LinkedIn, these could be assets to include on future job and college applications, much like the volunteer hours that P2's 501(c)(3) nonprofit community awarded. Programs like these and Discord University could encourage similar positive youth-led online communities, hopefully raising awareness among offline stakeholders like parents/guardians and schools. 

\subsubsection{Treating teen moderators as experts and stakeholders}
Our study showed how teen moderators and their communities can thrive when supported well, and these findings were only possible by talking to teen moderators directly. We urge more trust and safety researchers and practitioners to treat teen moderators as important stakeholders given their deep expertise and connection with other young users. It is no coincidence that past literature has positioned online community moderators as civic architects~\cite{matias2019civic}, and that modern debate about youth online safety evokes arguments about individual rights~\cite{ohlheiser_kosa_2024}. Youth subjects of these policies should be given a voice, and teen moderators can offer powerful information and context regarding these issues. 

Costanza-Chock argues that working with community leaders is critical to achieving true design justice~\cite{costanza2020design}, and HCI researchers have long advocated for community-based research that privileges the community members as experts~\cite{harrington2019deconstructing}. These philosophies should be applied to research with youth populations as well. Given their developmental stage, this type of work must be contextualized thoroughly, and Wiesnewski et al. have outlined the many benefits of approaching these problems from an autonomy- and resilience-based framework, especially in comparison with restriction- and surveillance-oriented measures that can actually backfire on this age group\cite{wisniewski2017parental,ghosh2018safety, wisniewski2015preventative, ghosh2018control}. Beyond researchers, more stakeholders should treat teens as direct stakeholders through mechanisms like youth advisory boards and participatory design~\cite{wisniewski2015resilience, park2023towards}. We have shown that teen moderators are especially valuable partners for trust and safety advocates in youth community-based platforms; we expect and hope that many teen moderators like our participants will someday become leaders of trust and safety in the future. 

\subsection{Why youth-led spaces are promising alternatives to other social media experiences}
\subsubsection{Curatable environments and community boundaries}
A core aspect of community-based platforms like Discord and Reddit that makes them different from SNS-based sites like Instagram and TikTok is their community membership boundaries. On Discord, members are either explicitly in or out of a specific community, and most social interactions happen in the context of these community units. In contrast, on Instagram, most interactions are based on user-to-user connections (i.e., follower lists). Community membership boundaries make social experiences much more controllable, which can be extremely valuable for stakeholders who want to curate healthy communities and influence the social interactions that teens might encounter online. This is precisely how the MMC --- designed by platform staff --- worked for teen moderators in our study. In the MMC, Discord admins were able to implement a highly selective application-based process by offering incentives like profile badges to cultivate certain types of values, cultures, and behavior. This approach focuses on \textit{proactively curating} a young user's environment instead of controlling their actions.

Another major benefit of community-based structures like Discord is that these community boundaries are easier to understand and program. To curate the community experience, participants in our study created bots and rules that reflect a community's values of leadership, justice, and conflict resolution. This curation process is less straightforward on network-based platforms like TikTok that rely on blackbox algorithms to deliver content to young users, which is one of the top concerns of legislators today~\cite{noauthor_global_2023}. Together, the bounded membership structure and transparent curation mechanisms make community-based platforms like Discord particularly promising for curating healthy and alternative youth-led social media experiences. 

\subsubsection{Encourages negotiating social relationships}
Negative experiences are inevitable in any social relationship, and bad actors will always exist online. A common mechanism found in network-based platforms like TikTok is that top-down moderation structures largely rely on banning, blocking, and filtering undesirable content to protect young users. While this is appropriate for egregiously harmful content, harm exists on a spectrum, and individuals do not automatically gain the resilience and skills to navigate these issues the moment they turn 18. Schoenebeck et al. have also shown that these approaches to harm resolution absolve offenders of their responsibilities and take away victims' chances for restoration~\cite{schoenebeck2021youth}. 

In contrast, an implicit feature of community-based platforms that can make them healthy alternatives for youth is that they are interest-based. In our findings, we surfaced that the voluntary participation, self-motivated nature, and genuine passion that teen moderators had in their communities were key to their success. When youth engage in online communities, it is typically because of an intrinsic motivation for a community's topic, whether it be cats, video games, or anime. Therefore, when conflicts inevitably occur, users have reason in community-based platforms to resolve and negotiate issues instead of blocking someone and moving on. These experiences practicing social contracts and navigating boundaries with peers are essential to adolescent identity development~\cite{davis2023technologys}. Furthermore, online communities are usually more forgiving to misunderstandings and mistakes since they are usually more pseudonymous and less attached to identity and profile than apps like Instagram. 

This discussion section covered only a select few of the implicit features that enabled the experiential learning benefits that teen moderators shared in our study. We offer these considerations to motivate similar future work about how different platform governance and moderation structures can influence healthier alternative experiences for youth in peer-led online communities.

\section{Limitations \& Future Work}
A primary limitation of our work is that our sample population is demographically homogeneous, and we lack visibility on MMC acceptance criteria. The participants in this study very much were a self-selected sample, so we do not claim that this is representative of all teen moderators. Future work could address this by providing more generalizable insights about Discord teen moderators based on a broader population. Due to the limited size and diversity of our sample, we are also unable to present a complete picture of the safety challenges that these moderators face. Our work indicates a serious need for longitudinal research on the emotional effects of moderation on teen volunteers, especially desensitization and burnout. We hope this inspires more research that maps the different types and amounts of support that distinct stakeholder groups are best suited to provide to keep teens safe online, especially teen moderators and their communities, since this balance between autonomy and safety is so key. Finally, we remark that while our discussion highlights Discord's platform structures as benefits, it is far from a perfect platform. Our work surfaces issues about its implicit hierarchy, feudalism~\cite{schneider2022feudalism}, and potential lack of demographic diversity, and future work should recruit more diverse youth populations and experiment with more alternative platform moderation and governance structures.

\section{Conclusion}
In this work, we contributed one of the first empirical studies with teen online community moderators. Through 13 interviews, we found that when youth online communities on Discord are well-supported, they can benefit greatly from community-driven moderation precisely \textit{because} it is community-driven. Moderators in our study gained valuable soft skills and professional experience, and they also facilitated spaces for healthy peer interactions. Their communities flourished as a direct result of their authentic leadership and values. We also surfaced some unique stakeholders including third-party businesses and support systems like the MMC, as well as offline adults and peers. Our findings show that the risks intrinsic to volunteer community-based moderation are evident and even more concerning for this population. We contextualize our findings to offer recommendations and opportunities to better support teen moderators and their communities. Our discussion reiterates that the self-led nature of these teen communities is central to their success and that stakeholders must recognize not to overstep what makes these spaces work in the first place. We also emphasize the importance of offering peer-led online communities as healthy alternative social online experiences for youth today. 

\begin{acks}
The authors would like to thank the participants and Discord staff who made this work possible. They would also like to thank the Yale Justice Collaboratory and the Trust \& Safety Professional Association (TSPA) for the opportunities to present and discuss the early stages of this work. This material is based upon work supported by the NSF CSGrad4US Fellowship under Grant No. G-1A-016. Any opinions, findings, and conclusions or recommendations expressed in this material are those of the authors and do not necessarily reflect the views of the National Science Foundation.

\end{acks}

\bibliographystyle{ACM-Reference-Format}
\bibliography{teendiscordmods}

\newpage
\appendix

\section{Appendix}

\subsection{Interview Protocol}
\textit{The following is a final version of the interview protocol which was revised through multiple iterations. Interviews were semi-structured, leaving space for follow-up questions. See Section 2 (Methods) for more details.}
\subsubsection{Warmup questions}
\begin{itemize}
    \item How many servers do you currently moderate?
    \item Can you tell me a little bit about each of them, including about how many members there are?
    \item Are there any other communities you used to moderate for that you no longer mod?
    \item Have you ever been a mod on other apps besides Discord?
\end{itemize}

\subsubsection{Becoming \& learning to be a mod}
\begin{itemize}
    \item When did you first become a moderator in any online community?
    \item Why did you want to become a moderator when you first started? Has that changed over time?
    \item If you’re ok with sharing, how old were you when you started?
    \item Were you asked about your age when you joined the team?
    \item Do you think the team cares about age?
    \item Did you know about Discord’s age policy?
    \item How did you learn what to do as a moderator in your first time modding?
    \item Was there anyone or anything in particular that you found really helpful?
    \item Are there any mentors or friends you learned from a lot?
\end{itemize}

\subsubsection{Personal growth \& philosophy}
\begin{itemize}
    \item Do you think you have grown as a person through moderating, or through Discord in general?
    \item What do you like about Discord or online communities in general? What do you like about them compared to other social media like Instagram or TikTok?
    \item Would you say most of the other people you talk to on Discord are around your age?
    \item Do you ever have different Discord accounts?
    \item Do you tend to use DMs more, or servers more?
    \item What about moderating specifically? What do you like about moderating?
    \item Do you see yourself being a mod in the long-term? Why or why not?
    \item Do you ever get burnt out or have issues with mod work life balance?
    \item How often do you check Discord? How do you balance it with school?
\end{itemize}

\subsubsection{Conflicts, leadership, \& governance}
\begin{itemize}
    \item Can you tell me about a recent interesting conflict or you had to resolve in any of your communities?
    \item What was your process and approach?
    \item How did you solve it?
    \item Is that how you typically resolve conflicts online?
    \item How much do you work with the other mods in your team?
    \item What are the different types of levels and roles for mods?
    \item Have you ever had a conflict between other moderators in your team? What did you do?
    \item How much do you interact with the rest of the server?
    \item What are some of the bots and tools your team uses for the server?
\end{itemize}

\subsubsection{Trust, safety, \& crises}
\begin{itemize}
    \item Have you ever interacted with any official people from Discord? What was the situation?
    \item What was the most serious situation or problem you ever had to deal with as a moderator? Did you feel like you knew what to do? Who did you go to for help?
    \item Have you heard of the term “child safety”? What does it mean to you? Have you ever experienced or witnessed any issues regarding that?
    \item What about “online safety” in general? What are the first things that come to mind when you hear those words?
    \item Do you feel like Discord’s Mod Academy material has stuff on that?
\end{itemize}

\subsubsection{Real life perceptions}
\begin{itemize}
    \item How much do you use Discord with your real life friends? More or less than people you don’t know? Do your friends from real life know about what you do as a moderator?
    \item Are there things you can talk to people on Discord about that you feel like you wouldn’t talk to people in real life about?
    \item Do your parents know that you moderate online? What do they think, or what do you think they would think if they knew?
    \item Has your school ever tried to incorporate Discord officially? Have they ever tried to ban it or other social media?
    \item Did you use Discord a lot during COVID lockdowns?
    \item Do you see yourself moderating in the long-term? Why or why not?
\end{itemize}

\subsubsection{Parting thoughts}
\begin{itemize}
    \item What are some things you think Discord could do better?
    \item What do you wish more people knew about Discord and moderating?
    \item Is there something else you’d want to share with me before we end the call?
\end{itemize}

\newpage

\subsection{Qualitative Codebook}
\textit{The following is a final version of the qualitative codebook after multiple iterations. See Section 2 (Methods) for more details.}
\begin{table}[h]
\resizebox{\textwidth}{!}{%
\begin{tabular}{lll}
\rowcolor[HTML]{434343} 
{\color[HTML]{FFFFFF} \textbf{Category}}               & {\color[HTML]{FFFFFF} \textbf{Code}}          & {\color[HTML]{FFFFFF} \textbf{Description}}                                                                                               \\
\rowcolor[HTML]{FFFFFF} 
\cellcolor[HTML]{D9D9D9}basic info                     & mod-experience                                & How long they have been a moderator                                                                                                       \\
\rowcolor[HTML]{F3F3F3} 
\cellcolor[HTML]{D9D9D9}basic info                     & about-communities                             & Information about the communities they moderate                                                                                           \\
\rowcolor[HTML]{FFFFFF} 
\cellcolor[HTML]{FFF2CC}skills learned from moderating & mod-career                                    & Mentions of moderation as part of their current or future career, job, or studies                                                         \\
\rowcolor[HTML]{F3F3F3} 
\cellcolor[HTML]{FFF2CC}skills learned from moderating & official-mediator                             & Interfacing between the community and some official business                                                                              \\
\rowcolor[HTML]{FFFFFF} 
\cellcolor[HTML]{FFF2CC}skills learned from moderating & socio-emotional                               & Mentions of how moderating or being in online communities influence social and emotional skills                                           \\
\rowcolor[HTML]{F3F3F3} 
\cellcolor[HTML]{FFF2CC}skills learned from moderating & work-professionalism                          & Descriptions of quasi-professional skills like navigating corporate culture, earning promotions, drafting resumes, and career networking. \\
\rowcolor[HTML]{FFFFFF} 
\cellcolor[HTML]{FFF2CC}skills learned from moderating & technical-skills                              & Descriptions of learning technical skills or using them for moderation                                                                    \\
\rowcolor[HTML]{F3F3F3} 
\cellcolor[HTML]{FFF2CC}skills learned from moderating & social capital                                & Making friends or connections online that can help them get ahead in other ways (e.g., job opportunities, other moderation gigs)          \\
\rowcolor[HTML]{FFFFFF} 
\cellcolor[HTML]{FFF2CC}skills learned from moderating & cross culture                                 & Descriptions of cultural exchanges or clashes, international/global context                                                               \\
\rowcolor[HTML]{F3F3F3} 
\cellcolor[HTML]{FFF2CC}skills learned from moderating & teamwork                                      & Working with/as a mod team, group decisionmaking                                                                                          \\
\rowcolor[HTML]{FFFFFF} 
\cellcolor[HTML]{FFF2CC}skills learned from moderating & communication-skills                          & Getting better at conveying and expressing themselves (getting better at understanding others falls under socio-emotional)                \\
\rowcolor[HTML]{F3F3F3} 
\cellcolor[HTML]{CFE2F3}moderator philosophies         & conflict-resolution                           & Resolving conflict within their communities                                                                                               \\
\rowcolor[HTML]{FFFFFF} 
\cellcolor[HTML]{CFE2F3}moderator philosophies         & governance-hierarchy                          & Descriptions of the community's structure and mod hierarchies                                                                             \\
\rowcolor[HTML]{F3F3F3} 
\cellcolor[HTML]{CFE2F3}moderator philosophies         & authority-tensions                            & Any mention of tensions, dislikes, and negative attitudes toward people with authority                                                    \\
\cellcolor[HTML]{CFE2F3}moderator philosophies         & \cellcolor[HTML]{F3F3F3}community-familiarity & \cellcolor[HTML]{FFFFFF}Importance of being part of or very familiar with the community                                                   \\
\rowcolor[HTML]{F3F3F3} 
\cellcolor[HTML]{CFE2F3}moderator philosophies         & procedural-justice                            & Fairness \& bias; showing awareness of how community perception influences justice                                                        \\
\rowcolor[HTML]{FFFFFF} 
\cellcolor[HTML]{CFE2F3}moderator philosophies         & restorative-justice                           & Methods that encourage healing victims, encouraging offenders to take responsibilities, or discourage future harm                         \\
\cellcolor[HTML]{CFE2F3}moderator philosophies         & \cellcolor[HTML]{FFFFFF}negative-examples     & \cellcolor[HTML]{F3F3F3}Negative examples of authority that they dislike or say they do NOT want to be like                               \\
\rowcolor[HTML]{FFFFFF} 
\cellcolor[HTML]{CFE2F3}moderator philosophies         & \textbf{earning-respect}                      & The idea that being a mod doesn't automatically grant respect                                                                             \\
\rowcolor[HTML]{F3F3F3} 
\cellcolor[HTML]{CFE2F3}moderator philosophies         & different-contexts                            & An implied understanding of how each community or person is different                                                                     \\
\rowcolor[HTML]{FFFFFF} 
\cellcolor[HTML]{EAD1DC}learning to mod                & discord-resources                             & Resources provided by Discord (Discord Mod Academy, trainings, Town Hall, etc)                                                            \\
\rowcolor[HTML]{F3F3F3} 
\cellcolor[HTML]{EAD1DC}learning to mod                & humans                                        & Descriptions of when moderating required some humanness, instinct, or nuance that a bot can't do                                          \\
\rowcolor[HTML]{FFFFFF} 
\cellcolor[HTML]{EAD1DC}learning to mod                & mentorship                                    & Mentors, mentorship; learning from someone with more experience (or teaching one with less)                                               \\
\rowcolor[HTML]{F3F3F3} 
\cellcolor[HTML]{EAD1DC}learning to mod                & mmc-dmd                                       & Any mentions of MMC or DMD specifically                                                                                                   \\
\rowcolor[HTML]{FFFFFF} 
\cellcolor[HTML]{EAD1DC}learning to mod                & meta-communities                              & Moderator Mentorship Community, Discord Moderator Discord, and other meta moderation communities                                          \\
\rowcolor[HTML]{F3F3F3} 
\cellcolor[HTML]{EAD1DC}learning to mod                & third-party                                   & Third-party resources for learning how to mod or Discord news                                                                             \\
\rowcolor[HTML]{FFFFFF} 
\cellcolor[HTML]{EAD1DC}learning to mod                & social-processing                             & Processing and discussing issues/problems with friends to make sense of and learn                                                         \\
\rowcolor[HTML]{F3F3F3} 
\cellcolor[HTML]{EAD1DC}learning to mod                & rules-guides                                  & Moderator rulebooks or guidebooks written with local/community-specific knowledge (does NOT include official Discord resources)           \\
\rowcolor[HTML]{FFFFFF} 
\cellcolor[HTML]{FCE5CD}motivations \& beginnings      & impact on others                              & Fulfillment from seeing impact of their actions on others                                                                                 \\
\cellcolor[HTML]{FCE5CD}motivations \& beginnings      & \cellcolor[HTML]{F3F3F3}standout-members      & Becoming a mod because they were exemplary community members                                                                              \\
\cellcolor[HTML]{FCE5CD}motivations \& beginnings      & community-affinity                            & Motivated to mod just because they really liked the community itself                                                                      \\
\cellcolor[HTML]{FCE5CD}motivations \& beginnings      & mod-applications                              & Formal applications to become a mod                                                                                                       \\
\cellcolor[HTML]{FCE5CD}motivations \& beginnings      & hobby-interests                               & Connecting over common hobbies or interests                                                                                               \\
\cellcolor[HTML]{FCE5CD}motivations \& beginnings      & influential-figures                           & Supporting or getting close to people of influence they look up to (streamers, artists)                                                   \\
\cellcolor[HTML]{D0E0E3}real life                      & irl-friends                                   & Friends, peers, other people their age, support networks in real life                                                                     \\
\cellcolor[HTML]{D0E0E3}real life                      & parents-school                                & Parents, school authorities; trusted adults                                                                                               \\
\cellcolor[HTML]{D9EAD3}recommendations                & age-maturity                                  & Descriptions of anything related to age (being asked age, perceptions of older or younger people, etc)                                    \\
\cellcolor[HTML]{D9EAD3}recommendations                & worklife-balance                              & Descriptions of moderating taking time from their other priorities. (Overlaps a lot with prioritization and mod-pipeline)                 \\
\cellcolor[HTML]{D9EAD3}recommendations                & platform-frustration                          & Being disappointed or frustrated with Discord as a platform or company                                                                    \\
\cellcolor[HTML]{D9EAD3}recommendations                & mod-pipeline                                  & When mods do different types of roles as they get older or gain more experience                                                           \\
\cellcolor[HTML]{D9EAD3}recommendations                & public-awareness                              & The need or desire to educate the general public about what moderating and online communities are                                         \\
\cellcolor[HTML]{D9EAD3}recommendations                & labor-pay                                     & Pay or paid opportunities, future career gigs                                                                                             \\
\cellcolor[HTML]{D9D2E9}trust \& safety                & bots-tools                                    & Bots, tools, technical skills, and programming related to moderation (overlaps with technical-skills)                                     \\
\cellcolor[HTML]{D9D2E9}trust \& safety                & crisis-emergency                              & Situations related to real life crises (child safety, suicide, doxxing)                                                                   \\
\cellcolor[HTML]{D9D2E9}trust \& safety                & harmful-content                               & Harmful content like gore, nudity, NSFW; attitudes toward it                                                                              \\
\cellcolor[HTML]{D9D2E9}trust \& safety                & legal-concerns                                & Concerns about how a legal issue or precaution affects moderation                                                                         \\
\cellcolor[HTML]{D9D2E9}trust \& safety                & \cellcolor[HTML]{F3F3F3}terms of service      & Mentions of ToS                                                                                                                           \\
\cellcolor[HTML]{D9D2E9}trust \& safety                & report-tools                                  & Built-in platform reporting tools                                                                                                         \\
\cellcolor[HTML]{D9D2E9}trust \& safety                & mod-harassment                                & Harassment or harm directed specifically at mods                                                                                          \\
\cellcolor[HTML]{D9D2E9}trust \& safety                & desensitization                               & Effects of exposure to harmful content                                                                                                    \\
\cellcolor[HTML]{D9D2E9}trust \& safety                & staff interactions                            & Interactions with Discord staff or admins                                                                                                 \\
\cellcolor[HTML]{E6B8AF}misc                           & social-media                                  & Thoughts on other platforms and social media like Instagram or TikTok                                                                     \\
\cellcolor[HTML]{E6B8AF}misc                           & misc                                          & Things that are interesting but don't have a category yet                                                                                 \\
\cellcolor[HTML]{E6B8AF}misc                           & \cellcolor[HTML]{FFFFFF}time-zones            & Any mention of time zones                                                                                                                 \\
\cellcolor[HTML]{E6B8AF}misc                           & \cellcolor[HTML]{F3F3F3}metaphors             & Metaphors of moderation                                                                                                                   \\
\cellcolor[HTML]{E6B8AF}misc                           & cross-platform                                & Whenever mods do stuff across platforms                                                                                                  
\end{tabular}%
}
\end{table}

\end{document}